\documentclass[%
 paper=a4, %
 twocolumn, %
 english, %
 DIV=16,
 headings=normal,
 headinclude=false,%
 footinclude=false,%
 ]{scrartcl}%

\usepackage[english]{babel}
\usepackage{ftnright}
\usepackage{amsmath}
\usepackage{graphicx}
\usepackage{amsmath}
\usepackage{bm}
\usepackage{siunitx}
\usepackage[colorinlistoftodos]{todonotes}
\usepackage{textcomp}

\addtokomafont{title}{\raggedright}
\addtokomafont{author}{\raggedright\normalsize}
\addtokomafont{date}{\raggedleft\normalsize}

\addtokomafont{captionlabel}{\sffamily\bfseries}%
\addtokomafont{caption}{\sffamily\footnotesize}

\setcapindent{0em}
\newcommand{\mos}{MoS\textsubscript{2}}
\newcommand{\El}{\ensuremath{E_0}}

\usepackage[backend=bibtex,style=nature, isbn=false]{biblatex}
\usepackage[colorlinks=true, allcolors=blue]{hyperref}
\newbibmacro{string+doi}[1]{%
  \iffieldundef{doi}{#1}{\href{http://dx.doi.org/\thefield{doi}}{#1}}}
\DeclareFieldFormat{title}{\usebibmacro{string+doi}{\mkbibemph{#1}}}
\DeclareFieldFormat[article]{title}{\usebibmacro{string+doi}{\mkbibquote{#1}}}

\begin{document}

\title{Measuring the Local Twist Angle and Layer Arrangement in Van der Waals Heterostructures}
\author{
  Tobias A. de Jong\textsuperscript{*,\textbf{1}},
  Johannes Jobst\textsuperscript{\textbf{ 1}},
  Hyobin Yoo\textsuperscript{\textbf{2}},
  Eugene E. Krasovskii\textsuperscript{\textbf{3,4,5}},
  Philip Kim\textsuperscript{\textbf{2}},\hfill\\
  Sense Jan van der Molen\textsuperscript{\textbf{1}}
}
  
\date{Submitted on 27 Apr 2018}

\maketitle  

\textbf{keywords:} { Low-energy electron microscopy, LEEM, van der Waals Materials, heterostacks, twist}\\[1cm]
\textbf{
The properties of Van der Waals heterostructures are determined by the twist angle and the interface between adjacent layers as well as their polytype and stacking. 
  Here we describe the use of spectroscopic Low Energy Electron Microscopy (LEEM) and micro Low Energy Electron Diffraction (\textmu LEED) methods to measure these properties locally. We present results on a MoS\textsubscript{2}/hBN heterostructure, but the methods are applicable to other materials.
  Diffraction spot analysis is used to assess the benefits of using hBN as a substrate.
  In addition, by making use of the broken rotational symmetry of the lattice, we determine the cleaving history of the MoS\textsubscript{2} flake, i.e., which layer stems from where in the bulk.}

\section{Introduction}

\let\thefootnote\relax\footnote{%
  \textsuperscript{1}\,Leiden Institute of Physics, Leiden University, Niels Bohrweg 2, P.O. Box 9504, NL-2300 RA Leiden, The Netherlands.\\
  \textsuperscript{2}\,Department of Physics, Harvard University, MA 02138, USA.\\
  \textsuperscript{3}\,Departamento de F\'isica de Materiales, Universidad del Pa\'is Vasco UPV/EHU, 20080 San Sebasti\'an/Donostia, Spain.\\
  \textsuperscript{4}\,Donostia International Physics Center (DIPC), Paseo de Manuel Lardizabal 4, 20018 San Sebasti\'an/Donostia, Spain.\\
  \textsuperscript{5}\,IKERBASQUE, Basque Foundation for Science, E-48013 Bilbao, Spain.\\
  * e-mail \href{mailto:jongt@physics.leidenuniv.nl}{jongt@physics.leidenuniv.nl}\\
  }
The list of materials that can be thinned down to single layers has been vastly augmented since the first isolation of graphene monolayers in 2004~\cite{novoselov2004}.
In the few layer limit these so-called Van der Waals (VdW) materials exhibit properties that are vastly different from their bulk counterparts, and they are hence interesting for fundamental research and applications alike~\cite{novoselov20162d}.
In particular the combination of different Van der Waals materials into heterostacks carry the potential for a wide range of applications~\cite{geim2013van}.

Mechanical exfoliation of single layers and their subsequent combination via stamping techniques makes it possible to create heterostacks of (almost) arbitrary layer arrangements. These methods have now advanced to the point that regular fabrication of multilayer heterostacks with sufficiently low defect density is commonplace.

The quality and properties of these heterostacks, however, are not only influenced by defects, but also critically depend on other factors such as the substrate, the crystallographic polytype of the layers and their relative orientation with respect to each other.
In particular, atomically flat substrates that do not perturb the electronic structure of the VdW stacks are desired, and consequently hexagonal boron nitride is widely used \cite{meric2013graphene,jobst2016quantifying}. 
The polytype of the different flakes, i.e. the different crystallographic configurations of the layers with respect to each other, determines many of the properties of VdW heterostacks. 
This point is particularly pronounced for transition metal dichalcogenides (TMDs) that can, for example, be semiconductors or metals depending on their polytype~\cite{nourbakhsh2016mos2}. 
Interlayer twist can cause stacking defects and strain, which can either result in a reduction of sample quality or in desired Moir\'e reconstructions. 
These reconstructions can strongly alter properties of the stacks such as their band structure~\cite{pamuk2017magnetic,yeh2016direct} or cause correlated electron effects culminating in the recent discovery of superconductivity in magic-angle bilayer graphene~\cite{cao2018unconventional}.

\begin{figure*}[htb!]%
  \includegraphics*[width=\textwidth]{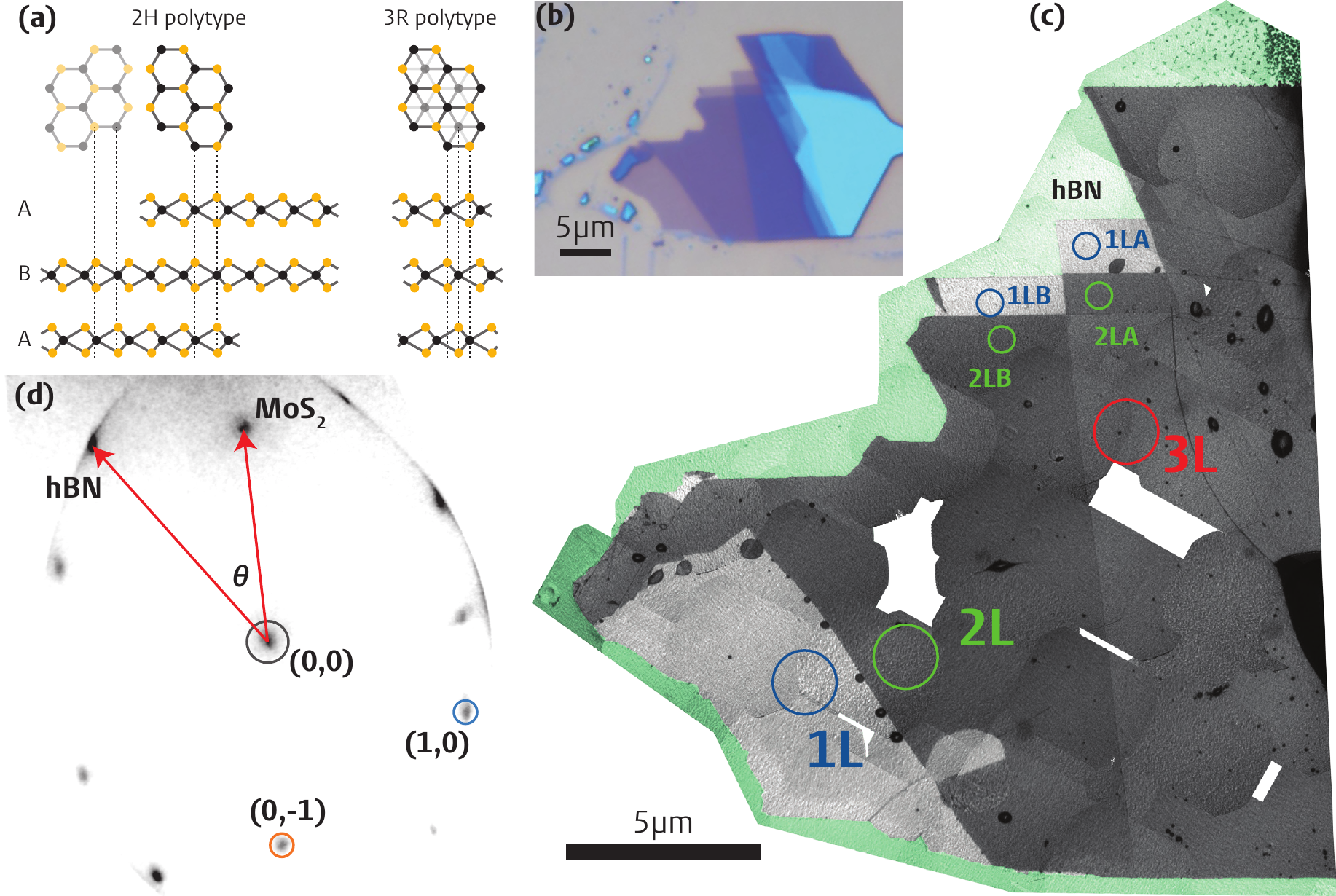}%
  \caption[]{%
  	\textbf{(a)} Top view and side view of the atomic structure of the 2H and 3R polytypes of \mos.
    \textbf{(b)} Optical microscope image of the MoS\textsubscript{2} flake on SiO before transfer, showing a clear layer contrast between different layer counts.
	\textbf{(c)} Overview of the MoS\textsubscript{2} flake composed of 90 bright-field LEEM micrographs. Dark areas and the bright area in the top left are MoS\textsubscript{2}, surroundings are hBN (colorized in green for clarity). Indicated are the areas studied with \textmu LEED measurements in Figure \ref{fig:mosmuleedlayer} and \ref{fig:mosmuleedab}. The number corresponds to the number of \mos\ layers. 
	\textbf{(d)} \textmu LEED measurements with both hBN as well as MoS\textsubscript{2} in the aperture. From this we determine the twist angle $\theta$ between the hBN and the MoS\textsubscript{2} as $\theta=29^\circ \pm 2^\circ$.
    }
    \label{fig:mosstacking}
\end{figure*}

In order to understand the properties of complex heterostructures, characterization techniques are needed to study the flatness of the interface, the relative rotation of the layers, and their polytype. 
Moreover, these techniques need to find the micrometer-sized heterostacks on millimeter-sized samples and simultaneously have sufficient lateral resolution to study details on the sub-micrometer length scale.

Typical characterization methods include, among others, optical microscopy, scanning electron microscopy (SEM) and atomic force microscopy (AFM) to obtain information about topography and thickness, Raman spectroscopy and angle-resolved photoemission spectroscopy (ARPES) for layer number and vibronic and electronic structure, respectively. 
Moreover, scanning tunneling microscopy (STM) and transmission electron microscopy (TEM) techniques allow advanced characterization down to the atomic level.
Although all these methods yield detailed insights into specific aspects of VdW heterostructures, they either can not simultaneously obtain information on flatness, layer number, rotation angle and polytype or need very specific sample preparation, e.g. free-standing samples for TEM investigations.

In this work, we demonstrate that all these parameters can conveniently be obtained within one setup using low-energy electron microscopy (LEEM) and diffraction (LEED).
We study a VdW heterostack of molybdenum disulfide (\mos) monolayer, bilayer and trilayer on bulk hexagonal boron nitride (hBN). 
This is a widely used material combination, but the methods demonstrated can be applied to virtually any heterostack on a large variety of substrates. 
We deduce flatness, layer number and polytype from spectroscopic LEEM measurements.

For the TMD \mos, there are three polytypes known, designated 1T, 2H and 3R. In this classification, the number reflects the number of layers in the unit cell, and the letter indicates whether the unit cell is hexagonal (H), rhombohedral (R) or trigonal (face-centered cubic, T).
The unit cell of the naturally most abundant, semiconducting 2H polytype consists of two layers of covalently bonded atoms [see Figure \ref{fig:mosstacking}(a)], which are weakly bonded by interlayer VdW force. 
In this polytype, subsequent layers are rotated $60^\circ$ with respect to each other.
Contrary to a simple hexagonal lattice such as graphene, which is sixfold symmetric, a single VdW layer is threefold rotation symmetric here (symmetry group $C_3$), i.e. a rotation over $120^\circ$ does not change the lattice.
Hence there are precisely two distinct types of layers in the 2H polytype, as a second rotation of $60^\circ$ yields a total rotation corresponding to the symmetry (total symmetry group $D_{3d}$).
This is in contrast to the 1T stacking, where all layers are identical, albeit different from the other polytypes, and the 3R stacking, where all layers have the same orientation, but lattice sites are shifted (as shown in Fig \ref{fig:mosstacking}(a)).
As the translation symmetry is purely defined by the shape of the unit cell, the reciprocal lattice is trigonal. Due to the threefold symmetry, the six first order diffraction spots are split in two equivalence classes, commonly denoted $K$ and $K'$.
In the following, we use the fact that these equivalence classes are visible in the observed LEED patterns of the samples studied and only consider the 2H and 3R polytypes.

\section{Methods}

\subsection{Experimental}

The VdW heterostack was fabricated through mechanical exfoliation and stamping: The MoS\textsubscript{2} flake was exfoliated using scotch tape onto silicon oxide for thickness determination and picked up again. Subsequently it was used to pick up a bottom hBN flake and transferred onto a silicon nitride substrate. 
Afterwards it was annealed at \SI{350}{\celsius} to remove polymer residue.

In LEEM, the sample is in ultra high vacuum and reheated to the same temperature. It is then illuminated with electrons of energies between $0$ and \SI{60}{\electronvolt}. 
This landing energy \El\ can be tuned precisely by changing a decelerating electric field between objective lens and sample~\cite{tromp2010new,tromp2013new}.
An image is formed from the reflected electrons, either in real space (LEEM image) or in reciprocal space (LEED pattern).
This combination of real space and $k$-space information and the ability to rapidly switch between the two, forms one of the strengths of LEEM instruments.

In this work, we use the SPECS P90 based ESCHER LEEM. 
This setup is aberration correcting, enabling a maximum resolution of \SI{1.4}{\nano\metre} ~\cite{schramm2011low}. 
In these experiments images are taken with a \SI{5}{\micro\metre} field of view. 
Larger images are created either in photoemission electron microscopy (PEEM) mode or by stitching multiple LEEM images together.

By scanning the electron energy \El\ and taking an image for each energy, spectroscopic data can be obtained. 
This can be done both in real space as well as in diffraction.
Spectroscopic \textmu LEED measurements are performed by limiting the illuminated area using an aperture and taking diffraction images for a range of landing energies.
This allows study of small areas of homogeneous layer number, to fingerprint the material and to determine layer number~\cite{hibino2008microscopic,de2016thickness}.
Spectra are in this work determined by averaging over an area around the respective diffraction spot for each energy. The data was corrected for detector effects and no smoothing of curves was performed.
The shape of the diffraction spots reveals additional information, e.g.\ the width of the central (0,0) spot (the specularly reflected electrons) is a measure for sample roughness~\cite{locatelli2010corrugation}. 
For determination of the spotwidth, a linecut along the maximum of the central peak and perpendicular to the dispersive direction of the prisms was taken, after the data was corrected for detector effects. 
The Full Width Half Maximum (FWHM) of a Gaussian fit to the top of this peak is then used as indicator for the sample flatness.

\subsection{Computational}
  The reflection spectra for specular and diffracted beams were calculated for different
  layer counts of 2H-MoS\textsubscript{2} using an ab initio theory of electron
  diffraction~\cite{flege2014intensity}. The calculations are performed with a
  full-potential linear augmented plane waves method with a self-consistent crystal
  potential obtained within the local density approximation as presented in
  Ref.\ \cite{Krasovskii1999}. The reflectivity spectra are obtained with the all-electron
  Bloch-wave-based scattering method of Ref.\ \cite{Krasovskii2004}, properly modified for
  stand-alone two-dimensional films of finite thickness~\cite{Nazarov2013}. The inelastic
  scattering was taken into account by introducing the optical potential: the imaginary
  potential $-iV_\text{i}$ is taken to be spatially constant over a finite slab (where the
  electron density is non-negligible) and to be zero in the two semi-infinite vacuum 
  half-spaces. We used the energy dependence $V_\text{i}(E)$ that was calculated in
  Ref.~\cite{siek2017} for a similar substance, WSe\textsubscript{2}, within the $GW$ approximation.

\section{Results and discussion}

\begin{figure}[htb]%
\includegraphics[width=\columnwidth]{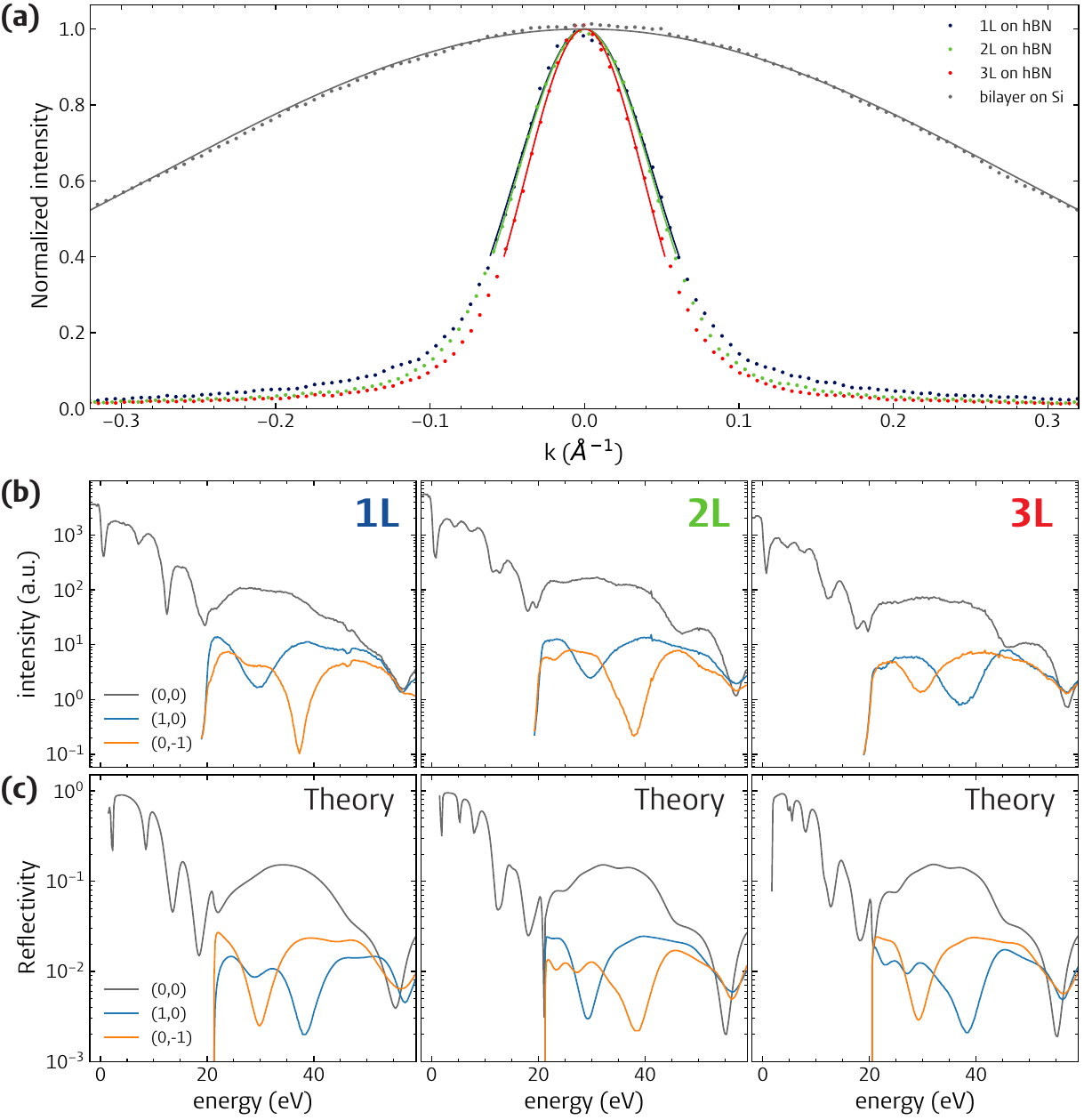}
	\caption{
		\textbf{(a)} Line profiles of the specular (0,0) \textmu LEED spot for monolayer, bilayer and triple layer areas of the \mos\ flake on hBN (as indicated in Figure \ref{fig:mosstacking}(c)) and for a bilayer on silicon (with native oxide). Gaussian fits to determine FWHM are shown as lines. All profiles are taken at \SI{32}{\electronvolt}, fitting procedure is described in Methods.
		\textbf{(b)} Comparison of the intensity of the specular \textmu LEED spot and of both refracted spots as a function of landing energy \El\ for the same areas as in (a). LEED spots used are indicated in Figure \ref{fig:mosstacking}(d).
\textbf{(c)} Theoretical calculations of low energy electron reflection for specular reflection and both equivalence classes of primary refracted spots for up to trilayer 2H-\mos, where each additional layer is added on top of the stack in the previous panel.
}
\label{fig:mosmuleedlayer}
\end{figure}

\subsection{Imaging and twist angle determination}

A thin \mos\ flake that contains monolayer, bilayer and trilayer areas was identified after exfoliation onto a Si/SiO$_2$ substrate. 
The layer number is clearly visible as different shades of purple in the optical microscopy image in Figure \ref{fig:mosstacking}(b). 
After stacking it onto a hBN flake and transferring the heterostructure to the final silicon nitride substrate (see methods above), PEEM mode is used in the LEEM setup to identify the flake. 
In the following, only LEEM data is discussed.

The entire heterostack is too big to be imaged in one LEEM field of view. Consequently, Figure \ref{fig:mosstacking}(c) shows an overview that is stitched together from 90 individual LEEM images (the white spots are areas of missing images). 
This LEEM overview reveals the features already visible in the optical image in much greater detail: a large monolayer (brightest), bilayer (darkest) and triple layer (intermediate gray) areas, as well as a region with smaller, rectangular monolayer and bilayer areas in the top right [as labeled in Figure \ref{fig:mosstacking}(c), where hBN is colored green for clarity]. 

In order to determine the angle between the \mos\ and the hBN crystal, we perform LEED experiments where both materials are illuminated by the electron beam. A LEED pattern taken on the edge of the flake is shown in Figure \ref{fig:mosstacking}(d).
Here, 50 images taken at even intervals between in the 50--\SI{60}{\electronvolt} range were averaged to enhance the signal-to-noise ratio and to capture all features. 
Two distinct hexagonal diffraction patterns are visible. 
The distance between the central (0,0) beam of specularly reflected electrons and the diffraction spots is inversely proportional to the lattice constant. Consequently, we can identify the diffraction spots further out as stemming from hBN and the ones further in stem from MoS\textsubscript{2}.
The twist angle $\theta$ between hBN and \mos\ diffraction spots, corresponds directly to the twist angle between the two materials. From Figure \ref{fig:mosstacking}(d) we determine $\theta = 29^\circ \pm 2^\circ$.

\subsection{\textmu LEED spectroscopy}

\textmu LEED measurements were performed for large areas of monolayer (`1L'), bilayer (`2L') and trilayer (`3L') \mos\ (the precise areas as limited by an illumination aperture are indicated in Figure \ref{fig:mosstacking}(c)). 
The FWHM of the $(0,\!0)$ spot at $\SI{32.0}{\electronvolt}$, indicative of flatness, was determined by fitting the spot profile [See Methods]. 
The profiles and resulting fits are shown in Figure \ref{fig:mosmuleedlayer}(a) for the mono- (blue), bi- (green), and trilayer (red) areas, resulting in FWHM values of respectively $\SI{0.107}{\per\angstrom}$, $\SI{0.104}{\per\angstrom}$ and $\SI{0.091}{\per\angstrom}$.
This is a factor of 6 lower than values for exfoliated MoS\textsubscript{2} on a silicon substrate (gray curve in Figure \ref{fig:mosmuleedlayer}(a), $\SI{0.66}{\per\angstrom}$). 
This large linewidth, due to the roughness of SiO\textsubscript{2}, confirms the findings of Yeh et al.~\cite{yeh2014probing}.
In contrast to their results, we find no significant broadening of the monolayer peak, indicating that the Van der Waals force between the atomically flat hBN and \mos\ effectively prevents buckling of the latter.
Combined, this reaffirms the significance of hBN as an atomically flat substrate for few layer Van der Waals devices.

Varying the energy and plotting the intensity of the specular and refracted beams as a function of energy results in the spectra shown in Figure \ref{fig:mosmuleedlayer}(b). 
Besides the lattice symmetry, lattice constants and angle between the layers, which we can all deduce from LEED images, these LEED spectra yield additional information. 
The specular spectra reflect the density of states of the material \cite{jobst2015nanoscale,jobst2016quantifying}. The diffracted beams, in addition, contain information about the symmetry properties of the material studied.
 
The threefold symmetry of the atomic lattice and inequivalence of $K$ and $K'$ for \mos\ is visible in experimental LEED data~\cite{yeh2014probing}: for some energies three of the six first order spots, at $120^\circ$ angles from each other, dim out.
Consequently, we choose one representative spot for $K$ and $K'$ each. They are denoted with their reciprocal lattice coordinates $(1,\!0)$ and $(0,\!-1)$, as indicated in Figure \ref{fig:mosstacking}(d).

The experimental spectra for the specular spot show a well-defined structure.
Differences between different layer counts are however subtle, the most prominent being the minimum at \SI{5.2}{\electronvolt} exhibited by multilayers, but not by the monolayer.
For comparison, we performed ab initio calculations of LEED reflectivity spectra for freestanding few layers 2H-\mos.
The results from these calculations match very well with the experimental data and are shown together in Figure \ref{fig:mosmuleedlayer}(c).

The calculated spectra show two classes of diffracted beams: the diffraction spots of the two equivalence classes have different intensities as a function of landing energy, with a pronounced dip at either $29$ or \SI{38}{\electronvolt}.
The experimental diffracted beams reproduce this behavior almost perfectly, with indeed minima at either $29$ or \SI{38}{\electronvolt}. 
A feature of the measurement not reproduced in the theoretical spectra is the increasing depth of the minimum at \SI{38}{\electronvolt} for decreasing layer number. 
We expect this to be due to the presence of the hBN substrate, which is not considered in the calculations.

The difference between spectra from the $(1,\!0)$ and $(0,\!-1)$ diffracted beams is a  result of the fact that the two layer types in 2H stacking are rotated 60 degrees with respect to each other (see Figure \ref{fig:mosstacking}(a)). Consequently, looking at the one or the other type of layer should interchange the behavior of the $K$ and $K'$ spots. 
In fact, as more layers are added on top in the calculations, the diffracted beams interchange behavior for each added layer, as expected from imaging layers of the different types. Therefore we conclude that the spectra from the first order beams are dominated by the top layer.

In the experimental curves (Figure \ref{fig:mosmuleedlayer}(b)), the spectra do not interchange from the monolayer to the bilayer case. 
This difference could have two causes as is visible from Figure \ref{fig:mosstacking}(a). The bilayer could either be of the 3R polytype, where both layers have the same orientation, or the layers are in 2H stacking but the second layer is added below the monolayer. 
The excellent match between experimental data and ab initio calculations does suggest 2H stacking for all areas. 
However, to fully rule out the presence of 3R stacking we perform additional experiments.

\begin{figure}[ht!]%
\includegraphics*[width=\linewidth]{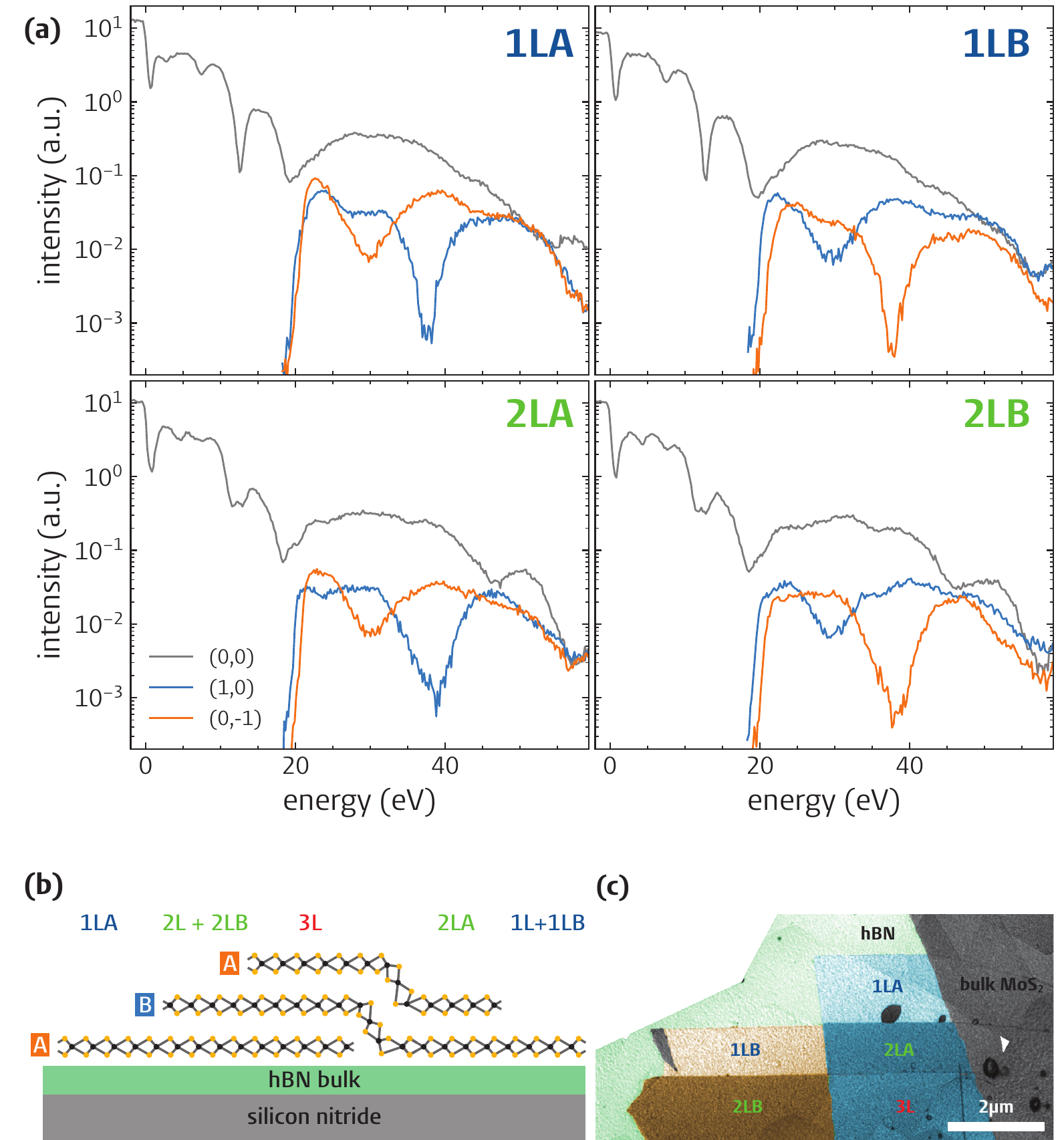}
\caption{%
	\textbf{(a)} \textmu LEED spectra of smaller neighboring areas, indicated in Figure \ref{fig:mosmuleedlayer}. For the different areas of the same layer count, equivalence classes of first order diffraction spots invert, indicating the top layers in the different areas stem from different layers in the bulk 2H crystal.
	\textbf{(b)} Cartoon showing a reconstruction of which areas stem from which layers in the original bulk crystal as concluded from (a) and and Figure \ref{fig:mosmuleedlayer}(b).
	\textbf{(c)} False color image indicating the top layer orientation as can be concluded from the spectra in (a).
	}
\label{fig:mosmuleedab}
\end{figure}

Further spectra, shown in Figure \ref{fig:mosmuleedab}(a), taken on smaller bilayer and monolayer areas (indicated in the top part of Figure \ref{fig:mosstacking}(c)), give additional evidence that the whole flake is 2H stacked:
As the flake is continuous, the spectrum inversion, where the diffraction spots switch equivalence class going from a monolayer to an adjacent bilayer area, proves the layers in the sample are 2H-stacked.
The fact that the asymmetry remains for the bilayer is thus not due to 3R stacking, but fully due to the low penetration depth of the low energy electrons, causing the spectra to be dominated by the topmost layer.

This notion now also helps to explain the diffracted curves for `1L' and `2L' in Figure \ref{fig:mosmuleedlayer}(b): contrary to the simulation the top layer here stays the same, the additional layer instead being added between the top layer and the substrate.

The fact that we can now determine the rotational type of the top layer, allows us to assign from which layer orientation in the bulk 2H-MoS\textsubscript{2} the top layer originates for different areas on the sample.
With the additional \textmu LEED measurements on smaller areas, we thus reconstruct the full cleaving history of the different layers of the MoS\textsubscript{2} flake in Figure \ref{fig:mosmuleedab}(b), determining for each boundary whether a layer is added/subtracted on top of the flake or on the bottom, information to the best of the authors' knowledge not measurable by any other technique.

\section{Conclusions}

In conclusion, we have shown the application of spectroscopic LEEM techniques to the characterization of Van der Waals heterostacks. 
The combination of real space imaging and local electron diffraction enables analysis of sample quality, stacking angle and polytype within one instrument, without the need for special substrates.
We conclude from the significantly reduced diffraction spot width compared to \mos\ layers on a silicon substrate, that the use of hBN as a substrate yields very high sample quality.
We compare experimental data with ab initio calculations, which allows us to locally distinguish the orientational type of the top layer and thus to conclude for each boundary of layer count whether a layer is added on top or on the bottom.

\section{Acknowledgement}
We thank Ruud M. Tromp for the fruitful discussions and useful advice. Furthermore we thank Marcel Hesselberth and Douwe Scholma for their indispensable technical support.
This work was supported by the Netherlands Organisation for Scientific Research (NWO/OCW) via the VENI grant (680-47-447, J.J.), as part of the Frontiers of Nanoscience program and
the Spanish Ministry of Economy and Competitiveness MINECO, Grant No. FIS2016-76617-P.

\printbibliography
\end{document}